\documentclass[twocolumn,runningheads,natbib]{svjour2}
\smartqed  
\usepackage{graphicx}

\journalname{Astrophysics and Space Science}
\begin{document}

\title{On the role of the current loss in radio pulsar
evolution
\thanks{This
work was partially supported by the Russian Foundation for Basic
Research (Grant no.~05-02-17700) and Dynasty fund. Elena Nokhrina
thanks the Conference Organizing Committee for the PPARC/Padova
University Grant, and the RFFR for the travel grant
(no.~06-02-26645).}}





\author{V.S.Beskin        \and
       E.E.Nokhrina
}


\institute{V.Beskin \at
             P.N.Lebedev Physical Institute, RAS,\\
             Leninsky pr. 53, Moscow, 119991 \\
             \email{beskin@lpi.ru}           
          \and
          E. Nokhrina \at
             Moscow Institute of Physics and Technology,\\
             Institutsky per. 9, Dolgoprudny, 141700 \\
             \email{nokhrinaelena@gmail.com}
}

\date{Received: date / Accepted: date}

\maketitle

\begin{abstract}
The aim of this article is to draw attention to the importance of
the electric current loss in the energy output of radio pulsars.
We remind that even the losses attributed to the magneto-dipole
radiation of a pulsar in vacuum can be written as a result of an
Ampere force action of the electric currens flowing over the
neutron star surface~\citep{Mich,BeskGurIst-93}. It is this force
that is responsible for the transfer of angular momentum of a
neutron star to an outgoing magneto-dipole wave. If a pulsar is
surrounded by plasma, and there is no longitudinal current in its
magnetosphere, there is no energy
loss~\citep{BeskGurIst-93,MPS-99}. It is the longitudinal current
closing within the pulsar polar cap that exerts the retardation
torque acting on the neutron star. This torque can be determined
if the structure of longitudinal current is known. Here we remind
of the solution by Beskin, Gurevitch \& Istomin (1993) and
discuss the validity of such an assumption. The behavior of the
recently observed "part-time job" pulsar B1931+24 can be
naturally explained within the model of current loss while the
magneto-dipole model faces difficulties.


\end{abstract}




\section{Magneto-dipole loss}
\label{sec:1}

The first idea to explain the energy loss of radio pulsars was to
consider the magneto-dipole radiation~\citep{Pacini-67}. Indeed,
the magneto-dipole formula gives for the radiation power
\begin{equation}
W_{{\mathrm md}}
=\frac{1}{6}\frac{B_{0}^{2}\Omega^{4}R^{6}}{c^{3}}\sin^{2}{\chi},
\label{MD-0}
\end{equation}
where $\chi$ is the angle between rotational and magnetic axis,
$R$ is a neutron star radius $\sim 10$ km, and $\Omega$ is a
pulsar angular velocity. This formula explains pulsar activity
and observed energy loss for expected large magnetic field near
the surface $B_{0}\sim 10^{12}$ Gs.

Let us recall that the physical reason of such energy loss is the
action of the torque exerted on the pulsar by the Ampere force of
the electric currens flowing over the neutron star surface
\citep{Ist-05}. The electric and magnetic fields in the outgoing
magneto-dipole wave in vacuum can be found by solving the wave
equations $\nabla^2\mathbf{B}+{\Omega^{2}}/{c^{2}}\mathbf{B}=0$
and $\nabla^2\mathbf{E}+{\Omega^{2}}/{c^{2}}\mathbf{E}=0$ with
the boundary conditions stated as the fields corresponding
components $\mathbf{E}_{\mathrm t}$ and $\mathbf{B}_{\mathrm n}$
being continuous through the neutron star surface. Inside the
star one can consider magnetic field as homogeneous, and find the
corresponding electric field using the frozen-in condition. As a
result, the full solution will give us the discontinuity of
$\{\mathbf{B}_{\mathrm t}\}$ and $\{\mathbf{E}_{\mathrm n}\}$
attributed to the surface charge $\sigma_{\mathrm s}$ and the
surface current $\displaystyle\mathbf{J}_{\mathrm
s}=\frac{c}{4\pi}[\mathbf{n},\{\mathbf{B}\}]$. The Ampere force
exerts the torque
\begin{equation}
\mathbf{K}=\frac{1}{c}\int{[\mathbf{r},[\mathbf{J}_{\mathrm
s},\mathbf{B}]]dS}
\end{equation}
on the neutron star. The energy loss of a pulsar due to this
torque is equal to (\ref{MD-0}). Thus, it is the surface current
that is responsible for the angular momentum transform from a
neutron star to an outgoing magneto-dipole
wave~\citep{Mich,BeskGurIst-93}.

A pulsar in vacuum loses its rotational energy due to angular
momentum transform to the electromagnetic wave at the rate given
by (\ref{MD-0}). However, this is not so if the pulsar
magnetosphere is filled with plasma and there is no longitudinal
current in the magnetosphere. As was shown by
\cite{BeskGurIst-93} and \cite{MPS-99}, in this case the Poynting
flux through the light cylinder is equal to zero. Indeed, as the
ideal conductivity condition is applicable not only inside the
neutron star but outside as well there is no magnetic field
discontinuity at the star surface. Consequently, there is no
Ampere force acting on a pulsar and, hence, there is no energy
loss. For zero longitudinal current the light cylinder is a
natural boundary of the pulsar magnetosphere.

\section{Current loss}

In this section we remind the exact solution for the surface
current within the polar cap presented in the monograph by
\cite{BeskGurIst-93}. As it was shown in the previous section,
the neutron star retardation is due to Ampere force ${\bf F}_{\rm
A} = {\bf J}_{\rm s} \times {\bf B}/c$. If the magnetosphere is
filled with plasma, the surface current ${\bf J}_{\rm s}$ is
flowing within magnetic polar cap only. This surface current
closes the volume longitudinal current in the magnetosphere and
the return current flowing along the separatrix between open and
closed field lines region.

In order to write the equation for the surface current, the
several assumptions must be made. We assume that the conductivity
of the pulsar surface is uniform, and the electric field ${\bf
E}_{\rm s}$ has a potential, so that the surface current can be
written as ${\bf J}_{\rm s} = {\bf {\nabla}}\xi'$. Using the
stationary continuity equation ${\rm div}{\bf J}=0$, where
$\partial J_{z}/\partial z$ is equal to the volume current
$i_{||}B_{0}$ flowing along the open field lines, one can obtain
\begin{equation}
\nabla^2 \xi' = -i_{||}B_{0}. \label{6}
\end{equation}

Making the substitution $x=\sin{\theta_m}$ and introducing the
non-dimensional potential $\xi=4\pi\xi'/B_{0}R^{2}\Omega$ and
current $i_{0}=-4\pi i_{||}/\Omega R^{2}$ we get
\begin{equation}
\left(1-x^{2}\right)\frac{\partial^{2}\xi}{\partial
x^{2}}+\frac{1-2x^{2}}{x}\frac{\partial\xi}{\partial
x}+\frac{1}{x^{2}}\frac{\partial^{2}\xi}{\partial\varphi_{m}^{2}}
= i_{0}(x,\varphi_{m}). \label{7}
\end{equation}
Here $\theta_m$ and $\varphi_{m}$ are polar and azimuth angles
with respect to magnetic axis.

Equation (\ref{7}) needs a boundary condition. This boundary
condition results from the proposition that there is no surface
current outside the magnetic polar cap. This means that
\begin{equation}
\xi\left[x_{0}(\varphi_{m}),\varphi_{m}\right] = {\rm const},
\label{a1}
\end{equation}
where $x_{0}(\varphi_{m})$ is the polar cap boundary. The
solution for the Dirichlet problem (\ref{7}, \ref{a1}) can be
obtained by the Fourier method. The jump in the potential
derivative at $x=x_{0}(\varphi_{m})$ gives us the current flowing
along the separatrix. As we see, it is defined uniquely by the
longitudinal current in the region of open field lines and by
condition that no longitudinal current can flow in the region of
the closed field lines.

For arbitrary inclination angle $\chi$ the current $i_{0}$ can be
written as a sum of its symmetric $i_{\rm S}$ and anti-symmetric
$i_{\rm A}$ components. The anti-symmetric current begins playing
the main role when the pulsar polar cap crosses the surface where
the Goldreich-Julian charge density $\rho_{\rm GJ}=-{\bf
\Omega}\cdot{\bf B}/2\pi c$ changes sign. This condition can be
written as
\begin{equation}
\chi=\frac{\pi}{2}-\sqrt{\frac{\Omega R}{c}}.
\end{equation}
For example, taking the Goldreich-Julian current density
\begin{equation}
i_{\rm GJ}(x,\varphi_{m}) \approx \cos{\chi} + \frac{3}{2} x
\cos\varphi_{m} \sin{\chi}=i_{\rm S}+i_{\rm A}x\cos\varphi_{m},
\label{8}
\end{equation}
we obtain the following solutions of the Dirichlet problem
(\ref{7})-(\ref{a1}) for the symmetric and anti-symmetric volume
currents respectively:
\begin{equation}
\xi_{\rm S}=\frac{i_{\rm S}}{4}x^{2},
\end{equation}
\begin{equation}
\xi_{\rm A}=\frac{i_{\rm
A}}{8}x(x^{2}-x_{0}^{2})\cos{\varphi_{m}}.
\end{equation}

The torque exerted by the surface current over the neuron star
can be written as
\begin{equation}
{\bf K} = \frac{1}{c}\int \left[{\bf r}, \left[{ \bf J}_{s},
\left({\bf B}_{0}\right)\right]\right]{\rm d}S, \label{9}
\end{equation}
where $\bf B_{0}$ is the dipole field near the neutron star
surface. Let us decompose the braking torque ${\bf K}$ over the
orthogonal system of unit vectors ${\bf e}_{m}$, ${\bf n}_{1}$,
and ${\bf n}_{2}$. Here ${\bf e}_{m}$ is a unit vector along the
magnetic moment; vector ${\bf n}_{1}$ is perpendicular to the
magnetic moment and lies in the plane of the magnetic moment and
the rotational axis; vector ${\bf n}_{2}$ complements these to
the right-hand triple:
\begin{equation}
{\bf K}=K_{||}{\bf e}_{m}+K_{\bot}{\bf n}_{1} +K_{\dag}{\bf
n}_{2}. \label{10}
\end{equation}
$K_{\dag}$ plays no role in Euler equations that describe the
rotational dynamics of the decelerating neutron star. As a result
we have~\citep{BeskGurIst-93}:
\begin{equation}
K_{||}= -\frac{B_{0}^{2}\Omega R^{4}}{2\pi c} \int_{0}^{2\pi}{\rm
d}\varphi_{m}\int_{0}^{x_{0}(\varphi_{m})}{\rm d}x\,
x^{2}\sqrt{1-x^{2}}\frac{\partial\xi}{\partial\varphi_{m}},
\label{11}
\end{equation}
\begin{equation}
K_{\perp}=K_{1}+K_{2}, \label{12}
\end{equation}
\begin{eqnarray}
K_{1} & = & \frac{B_{0}^{2}\Omega R^{4}}{2\pi
c}\int_{0}^{2\pi}{\rm d}\varphi_{m}
\int_{0}^{x_{0}(\varphi_{m})}{\rm d}x A,
\label{13} \\
K_{2} & = & \frac{B_{0}^{2}\Omega R^{4}}{2\pi
c}\int_{0}^{2\pi}{\rm d}\varphi_{m}
\int_{0}^{x_{0}(\varphi_{m})}{\rm d}x\,
x^{3}\cos{\varphi_{m}}\frac{\partial\xi}{\partial x}, \label{14}
\end{eqnarray}
where $\displaystyle A=x\cos{\varphi_{m}}{\partial\xi}/{\partial
x} -\sin{\varphi_{m}}{\partial\xi}/{\partial\varphi_{m}}$.

The leading perpendicular torque component $K_{1}$ is equal to
zero equivalently for arbitrary shape of the polar cap due to the
boundary condition (\ref{a1}). The values $K_{||}$ and $K_{\perp}$
can be written as
\begin{eqnarray}
K_{||} & = &
\frac{B_{0}^{2}\Omega^{3}R^{6}}{c^{3}}\left[-c_{||}i_{\rm S}
+\mu_{||}\left(\frac{\Omega R}{c}\right)^{1/2}i_{\rm A}\right],
\label{16} \\
K_{\perp} & = & \frac{B_{0}^{2}\Omega^{3}R^{6}}{c^{3}}
\left[\mu_{\perp}\left(\frac{\Omega R}{c}\right)^{1/2}i_{\rm S}
+c_{\perp}\left(\frac{\Omega R}{c}\right)i_{\rm A}\right].
\label{17}
\end{eqnarray}
Here the coefficients $\mu_{||}$ and $\mu_{\perp}$ depending on
the shape of the polar cap are much less than unity, and the
coefficients $c_{||}$ and $c_{\perp}\sim 1$.

We can now find the derivatives of the angular velocity
$\dot{\Omega}$ and of the inclination angle $\dot{\chi}$ of a
neutron star through the Euler dynamics equations:
\begin{eqnarray}
J_r\frac{{\rm d}\Omega}{{\rm d}t} & = &
K_{||}\cos{\chi}+K_{\perp}\sin{\chi},
\label{18} \\
J_r\Omega\frac{{\rm d}\chi}{{\rm d}t} & = &
K_{\perp}\cos{\chi}-K_{||}\sin{\chi}. \label{19}
\end{eqnarray}
For the inclination angles $\chi$ not too close to $90^{\circ}$
(i.e., for $\cos\chi> (\Omega R/c)^{1/2}$), when the
anti-symmetric current plays no role in the neutron star dynamics,
we obtain
\begin{eqnarray}
\label{Losses1} \frac{{\rm d}\Omega}{{\rm d}t} & = &
-c_{||}\frac{B_{0}^{2}\Omega^{3}R^{6}}{J_r c^{3}}i_{\rm
S}\cos{\chi},
\label{20} \\
\frac{{\rm d}\chi}{{\rm d}t} & = &
\phantom{-}c_{||}\frac{B_{0}^{2}\Omega^{2}R^{6}}{J_r c^{3}}i_{\rm
S}\sin{\chi}. \label{21}
\end{eqnarray}

As a result, for homogeneous current density within open magnetic
field lines region $i_{\mathrm S} = j_{||}/j_{\mathrm GJ} =$ const
where $j_{\mathrm GJ}= c \rho_{\mathrm GJ}$ we have
\begin{equation}
W_{\mathrm c} = \frac{f_*^2}{4} \,
\frac{B_{0}^{2}\Omega^{4}R^{6}}{c^{3}} i_{\mathrm S}\cos{\chi}.
\label{c-0}
\end{equation}
Here $f_{*}$ is the non-dimensional area of a pulsar polar cap:
$S_{\rm cap}=f_{*}\pi(\Omega R/c)$. It depends on the structure
of the magnetic field near the light cylinder. For a pure dipole
magnetic field (and aligned rotator) $f_* = 1$, and for a
magnetosphere containing no longitudinal currents $f_*$ changes
from $1.592$ for the aligned rotator, $\chi=0^\circ$ \citep{Mich},
to $1.96$ for an orthogonal rotator, $\chi = 90^{\circ}$
\citep{BeskGurIst-93}. Recent numerical calculations for an
axisymmetric magnetosphere with non-zero longitudinal electric
current give $f_* \approx 1.23$ --
$1.27$~\citep{Gruzinov-05,Komissarov-06,Timokhin-06}. If the
singular point separating open and close field lines can be
located inside the light cylinder, the value $f_*$ can be even
$\gg 1$. As the Goldreich-Julian charge density near the polar
cap is propotrional to $\cos\chi$, one can write
\begin{equation}
W_{\mathrm c} \approx \frac{f_*^2}{4} \,
\frac{B_{0}^{2}\Omega^{4}R^{6}}{c^{3}} \cos^2{\chi}. \label{c-1}
\end{equation}

On the other hand for $\chi \approx 90^{\circ}$ when the
anti-symmetric current plays the leading role we obtain
\begin{equation}
W_{\mathrm c} \approx \frac{B_{0}^{2}\Omega^{5}R^{7}}{c^{4}}.
\label{c-2}
\end{equation}
As we see, the energy loss of the orthogonal rotator are $\Omega
R/c$ times smaller than of the aligned rotator.

\section{PSR B1931+24}

The recent discovery of the "part-time job" pulsars PSR B1931+24
\citep{Lyn-06} with $\dot\Omega_{\rm on}/\dot\Omega_{\rm off}
\approx 1.5$ shows that the current loss is indeed playing an
important role in the pulsar energy loss. If we assume that in
the on-state the energy loss is connected with the current loss
only and in the off-state with the magneto-dipole radiation (in
which case the magnetosphere must be not filled with plasma) we
get
\begin{equation}
\frac{\dot\Omega_{\rm on}}{\dot\Omega_{\rm off}} = \frac{3
f_*^2}{2} \, {\rm cot}^2\chi.
\end{equation}
It gives $\chi \approx 60^{\circ}$. On the other hand,
%
if we assume the Spitkovsky's relation for the on-state energy
loss \citep{Spi-06}
\begin{equation}
W_{\rm tot} = \frac{1}{4} \, \frac{B_0^2 \Omega^4 R^6}{c^3} (1 +
\sin^2\chi),
\end{equation}
we obtain
\begin{equation}
\frac{\dot\Omega_{\rm on}}{\dot\Omega_{\rm off}} = \frac{3}{2} \,
\frac{(1+ \sin^2\chi)}{\sin^2\chi}.
\end{equation}
Clear, this ratio cannot be equal to 1.5 for any inclination
angle. This discrepancy can be connected with that fact that all
the numerical calculations produced recently contain no
restriction on the longitudinal electric current magnitude. As a
result, current density can be much larger than Goldreich-Julian
one.

\section{On the magnitude of a surface current}

As we have shown, the current loss plays the major role in the
pulsar dynamics. In particular, the behaviour of the pulsar
B1931+24 can be naturally explained within this model. The
current loss model have some important consequences:
\begin{enumerate}
\item the energy loss of an orthogonal rotator is $\Omega R/c$
times smaller than for an aligned rotator. This is connected with
the boundary condition (\ref{a1}) that leads to almost full
screening of the toroidal magnetic field in the open field lines
region (see \cite{BesNok-04});
\item consequently, during its evolution a pulsar inclination
angle tends to $\pi/2$ where energy loss is minimal.
\end{enumerate}
On the other hand, it is known for the Michel's monopole solution
that in order to have the MHD flow up to infinity, the toroidal
magnetic field must be of the same order as the poloidal electric
field on the light cylinder. If the longitudinal current $j_{||}$
does not exceed by $(\Omega R/c)^{-1/2}$ times the respective
Goldreich-Julian current density (for the typical pulsars this
factor approach the value of $10^2$), the light surface
$|{\mathbf{E}}| = |\mathbf{B}|$ for the orthogonal rotator must
be located in the vicinity of the light cylinder. In this case
the effective energy conversion and the current closure is to
take place in the boundary layer near the light surface
\citep{BeskGurIst-93,ChiLiBeg-98,BesRaf-00}.

In order for these results being not true (for example, in order
for the light surface being removed to infinity)
there must be a sufficient change in the current density value in
the inner gap. We should emphasize that for the model with free
particle escape it is hard to support the current different than
the Goldriech-Julian current. Indeed, since $\rho_{\rm GJ}$ is
the particle density needed to screen the longitudinal electric
field, the value for the current must be close $j_{\rm
GJ}=c\rho_{\rm GJ}$. In order to change this value significantly
we must support the plasma inflow in the inner gap region
\citep{Lyu-92}. For example, these particles can be produced in
the outer gap. But for different poloidal field configuration it
is obvious that the major number of field lines intersect the
outer gap region outside the Alfvenic surface: as it was shown for
several field configurations by \cite{BesKuzRaf-98} and
\cite{BesNok-06}, inside the fast magnetosonic surface the flow
remains still highly magnetized. Thus, the deviation of current
lines from the field lines is negligible. On the other hand,
magnetized plasma can intersect the Alfvenic surface outwards only
(see \cite{Bes-06} for more detail). Thus, the outer gap can not
significantly affect the current in the vicinity of the polar cap.

Finally, it is necessary to stress that the recent numerical
calculations by \cite{Gruzinov-05}, \cite{Komissarov-06},
\cite{Timokhin-06}, \cite{Spi-06} do not include into
consideration the condition that the longitudinal current density
must be close to $j_{\rm GJ}$. In all these works the authors
assume that there may be any current flowing through the cascade
region. However, if this is not so, and the current is indeed
close to the Goldreich-Julian current, the structure of a
magnetosphere may be different from the one obtained in the
numerical simulations.

\section{Conclusions}

As we have seen, the current loss connected with the longitudinal
current flowing in the magnetosphere plays the main role in pulsar
dynamics, and recent observations of "part-time job" pulsar
supports this point. This evolution includes not only a neutron
star retardation but also the sufficient change in the angle
between magnetic and spin axis. We have seen as well that the
model of current loss depends crucially on the distribution of the
electric current and its value in the inner gap. For current loss
model with $j_{\rm GJ}$ the inclination angle grows with time so a
pulsar tends to be an orthogonal rotator. In this case the energy
loss is to be $\Omega R/c$ times smaller than for the aligned
rotator. As a consequence, the light surface must be located in
the very vicinity of the light cylinder. On the other hand, to
realize the homogeneous MHD outflow up to infinity for the
orthogonal rotator the current density in the open field line
region is to be much larger than $j_{\rm GJ}$.

\begin{acknowledgements}
The authors thank A.V.Gurevich and Ya.N.Istomin for the useful
discussions.

\end{acknowledgements}



\end{document}